\documentclass[a4paper, 10pt, unnumberedsections, twoside]{LTJournalArticle}

\usepackage{lipsum}
\usepackage[acronym]{glossaries}
\usepackage[detect-all=true,per-mode=symbol,per-symbol =/]{siunitx}
\usepackage[capitalise]{cleveref}
\usepackage{tablefootnote}
\usepackage{threeparttable}
\usepackage{booktabs}
\usepackage{xcolor}
\usepackage{makecell}
\usepackage{subcaption}

\DeclareSIUnit{\x}{\!\ensuremath{\times}}
\DeclareSIUnit\bit{b}
\DeclareSIUnit\gateeq{GE}
\runninghead{}
\footertext{\textit{RISC-V Summit Europe, Paris, 12-15 May 2025}}
\title{Croc: An End-to-End Open-Source Extensible RISC-V MCU Platform to Democratize Silicon}

\author{%
    Phillippe Sauter\textsuperscript{1}\thanks{Corresponding author: \href{mailto:phsauter@iis.ee.ethz.ch}{\tt phsauter@iis.ee.ethz.ch}}, \
    Thomas Benz\textsuperscript{1}, \
    Paul Scheffler\textsuperscript{1}, \
    Hannah Pochert\textsuperscript{1}, \
    Luisa Wüthrich\textsuperscript{1}, \\
    Martin Povišer, \
    Beat Muheim\textsuperscript{1}, \
    Frank K. Gürkaynak\textsuperscript{1}, \
    Luca Benini\textsuperscript{1,2}
}

\newacronym{asic}{ASIC}{application-specific integrated circuit}
\newacronym{eda}{EDA}{electronic design automation}
\newacronym{fpga}{FPGA}{field-programmable gate array}
\newacronym{ip}{IP}{intellectual propertie} %
\newacronym{os}{OS}{open-source}
\newacronym{pdk}{PDK}{process design kit}
\newacronym{rtl}{RTL}{register transfer level}
\newacronym{soc}{SoC}{system-on-chip}
\newacronym{uc}{MCU}{microcontroller}
\newacronym{vlsi}{VLSI}{very-large-scale integration}
\newacronym{isa}{ISA}{instruction set architecture}
\newacronym{sv}{SV}{SystemVerilog}

\newcommand{\x}{$\times$}
\newcommand{\riscv}{\mbox{RISC-V}}

\date{
    \vspace{-0.32em}
    \footnotesize\textsuperscript{\textbf{1}}Integrated Systems Laboratory, ETH Zurich \\
    \footnotesize\textsuperscript{\textbf{2}}Department of Electrical, Electronic, and
    Information Engineering, University of Bologna
    \vspace{-0.32em}
}

\renewcommand{\maketitlehookd}{%
	\begin{abstract}
		\noindent 
        Ensuring a continuous and growing influx of skilled chip designers and a smooth path from education to innovation are key goals for several national and international "Chips Acts".
        Silicon democratization can greatly benefit from end-to-end (from silicon technology to software) free and \gls{os} platforms.
        We present Croc, an extensible {\riscv} microcontroller platform explicitly targeted at hands-on teaching and innovation.
        Croc features a streamlined \gls{os} synthesis and an end-to-end \gls{os} implementation flow, ensuring full, unconstrained access to the design, the design automation tools, and the implementation technology.
        Croc uses the industry-proven, open-source CVE2 core, implementing the RV32I(EMC) \gls{isa}, enabling students to define and implement their own \gls{isa} extensions.
        MLEM, a tapeout of Croc in IHP's open \SI{130}{\nano\metre} node completed in eight weeks by a team of just two students, demonstrates the platform's viability for hands-on teaching in schools, universities, or even on a self-education path.
        In spring 2025, ETH Zurich will utilize Croc for its curricular VLSI class, involving up to 80 students, producing up to 40 \gls{os} application-specific integrated circuit layouts, and completing up to five student-led system-on-chip tapeouts.
        The lecture notes and exercises are already available under a Creative Commons license. 
	\end{abstract}
}

\begin{document}

\maketitle

\glsresetall

\section{Introduction}

Silicon democratization is a key objective of large \emph{Chips Acts} launched worldwide in response to the skill shortage in \gls{vlsi} design~\cite{2024europeanchipsact, 2024chipsandscienceact, 2024indiainjects}.
An \gls{os} hardware design approach~\cite{2024importance, 2024edu4chips} is key to achieving silicon democratization, as it enables large cohorts of students to gain access to technology \glspl{pdk}, \gls{eda} tools, and design \glspl{ip}, which are required assets for chip design.
The \gls{os} approach can facilitate innovation, as students with hands-on experience can easily transition from school to industry or start-ups.

The current status quo for \gls{vlsi} courses is to focus on teaching a theoretical understanding of the design and fabrication process~\cite{2024edu4chips}.
Those featuring practical lab modules often only \emph{emulate} the \gls{asic} design flow using \glspl{fpga}~\cite{2024edu4chips}.

\Gls{os} eases a more aggressive hands-on approach to teaching \gls{vlsi}.
\emph{Edu4Chip}~\cite{2024edu4chips} targets a strong multinational undergraduate-level chip design course, teaching the theoretical background and offering hands-on experience for $\geq$~250 students annually. 
This course uses \emph{Didactic-SoC}~\cite{2024edu4chips}, a chip platform created with \emph{Kactus2} using \emph{IP-XACT} descriptions.
It features a \emph{staff} section containing a {\riscv} \gls{soc} and multiple \emph{student} sections.
At the time of writing, their platform neither features a synthesis nor implementation flow and heavily uses generated code, steepening the learning curve for students.

With \emph{Croc}~\cite{pulpplatform2024croc}, we provide a more mature platform based on production-ready, industry-proven \glspl{ip}, featuring a simple and proven physical implementation flow with supplementary exercises and lecture notes.
Students gain hands-on experience on a mature set of \glspl{ip} that helps bridge the gap between teaching and industry. 
By minimizing the size of the {\riscv} \gls{uc}, we allow students to work on their own \gls{asic} end-to-end, giving them complete control over the entire design and implementation. 
To reduce the barrier to entry, we rely solely on well-documented, parameterized, and silicon-proven \emph{\gls{sv}} code and a streamlined implementation flow.

In November 2024, two bachelor students successfully taped out \emph{MLEM}~\cite{asic2024mlem} to demonstrate Croc's implementation feasibility in \emph{IHP}'s open 130-\si{nm} node~\cite{herman2024reflections}.

Starting in spring 2025, ETH Zurich's \gls{asic} design course, \emph{VLSI2}, switches to an end-to-end \gls{os} flow using Croc.
All teaching material is released under a \emph{Creative Commons} license.~\footnote{\url{vlsi.ethz.ch}}

We present the following contributions:
\begin{itemize}
    \item An education-focused, extensible \gls{asic} platform featuring a minimal {\riscv} \gls{uc} implemented with silicon-proven production-ready \gls{sv} \glspl{ip}.
    \item A streamlined and documented \gls{os} \gls{eda} synthesis and implementation flow based on our work shown at IWLS 2024~\cite{sauter2024insights} and our newly developed \emph{yosys-slang}~\footnote{\url{github.com/povik/yosys-slang}} \gls{sv} frontend for \emph{Yosys}. 
    \item A successful end-to-end open IHP \SI{130}{\nano\metre} demonstrator tapeout in November of 2024, silicon-proving the \gls{soc} platform for educational courses.
\end{itemize}

\section{Croc Platform}

\begin{figure}[t]
    \begin{subcaptionblock}{0.285\linewidth}%
        \centering%
        \includegraphics[width=\linewidth]{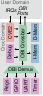}%
        \vspace{-0.3em}%
        \caption{}%
        \label{fig:croc-arch}%
    \end{subcaptionblock}\hfill
    \begin{subcaptionblock}{0.65\linewidth}
        \centering%
        \includegraphics[width=\linewidth]{fig-02.png}%
        \vspace{-0.3em}%
        \caption{}%
        \label{fig:mlem-render}%
    \end{subcaptionblock}\hfill
    \vspace{-0.9em}%
    \caption{(a) Architecture Croc, (b) MLEM's layout.}
    \label{fig:croc-mlem}
\end{figure}

The supporting \emph{Croc} domain contains an \gls{os}, industry-maintained, production-ready, Ibex-based~\cite{lowrisc2015ibex}
 \emph{CVE2}~\cite{openhw2022cve2} {\riscv} core, a minimal set of peripherals, and an \emph{OBI}~\cite{silicon2020obi1} crossbar.
The Croc domain primarily aids the students in implementing, debugging, and verifying their designs.
Students may modify it by, for example, by implementing custom \gls{isa} extensions or modifying the interconnect.
Croc features a single-cycle tightly coupled interconnect and two {SRAM} banks, allowing {CVE2} to achieve its ideal performance of one instruction per cycle.
The \emph{user domain} provides an interface for loosely coupled accelerators, peripherals, or experimental {\riscv} cores; see \Cref{fig:croc-arch}.
Croc is available ready to use from a single repository~\footnote{\url{github.com/pulp-platform/croc}} with the \gls{rtl} description, software setup, and documentation.

We target IHP's \SI{130}{\nano\metre} open \gls{pdk}~\cite{herman2024reflections} with a fully \gls{os} design flow using \emph{Yosys} for synthesis, \emph{OpenRoad} to implement the backend, and \emph{Verilator} for \gls{rtl} simulation.
Croc's flow is a streamlined version of our previous \emph{IWLS 2024}~\cite{sauter2024insights} flow. 
Most notably, we replaced the complex \gls{rtl} preprocessing step with our newly developed \emph{slang-based} \gls{sv} frontend for Yosys~\footnote{\url{github.com/povik/yosys-slang}}.
The \emph{IIC-OSIC-TOOLS}~\cite{iic-jku2024iic-osic-tools} container provides the \gls{os} \gls{eda} tools and the \gls{pdk}.
We provide an FPGA flow targeting \emph{Digilent}'s \emph{Genesys 2} to support accessible, low-cost verification and emulation.

\section{MLEM Student Tapeout}

In their Bachelor's thesis, two students took the lead in extending Croc with an optimized \emph{UART} peripheral and a \emph{NeoPixel}~\cite{burgess2013adafruit} controller, culminating in the successful tapeout of our demonstrator chip MLEM, presented in \Cref{fig:mlem-render}. 
Using a predecessor of the new VLSI2 exercises, they independently designed, implemented, and verified their designs, completed the physical design flow in eight weeks, and contributed their experiences to our shared knowledge base.

MLEM measures \SI{5}{\milli\metre\squared} and has a design complexity of \SI{350}{kGE} at a global density of \SI{56}{\percent}.
MLEM can be implemented on a \emph{6$^{th}$ Generation Intel Core {i7}} machine in less than one hour with a memory footprint of less than eight \si{\gibi\byte}.
Of the 48 I/O pads, 12 are used by the Croc domain. 
The remaining 36 pins carry NeoPixel, UART, and 26 GPIOs.
Under typical conditions, the design achieves a top clock speed of \SI{80}{\mega\hertz} (58 logic levels) at a core voltage of \SI{1.2}{\volt}.

\section{Conclusion and Outlook}

We present Croc, an education-focused, extensible end-to-end \gls{os} {\riscv} \gls{uc} platform built around mature, silicon-proven \gls{sv} \glspl{ip} used in multiple commercial projects and industry-supported repositories.
Croc thus bridges the gap between teaching and industry, allowing the students to use this platform as a starting kit to develop commercial {\riscv} \gls{uc} products, e.g., for security, control, or edge machine learning applications.
Croc comes bundled with a comprehensible \gls{os} \gls{eda} synthesis and implementation flow together with educational material to build physical implementation experience.
We successfully taped out MLEM, establishing Croc as the \gls{asic} platform for our VLSI2 course.
In 2025, VLSI2 will educate approximately 80 students at ETH alone, producing up to 40 \gls{os} \gls{asic} layouts and up to five \gls{soc} tapeouts using IHP's \SI{130}{\nano\metre} \gls{pdk}.

\vspace{-0.2cm}
\noindent\paragraph{\textbf{Acknowledgment}:} %
This work was supported through the SwissChips Initiative.
Silicon is funded by the German BMBF project FMD-QNC (16ME0831).

\printbibliography

\end{document}